\documentclass[a4paper]{jpconf}

\usepackage{graphicx}

\usepackage{epsfig}
\usepackage{wrapfig}
\usepackage{amsmath}
\usepackage{amssymb}
\usepackage{multicol,sectionbox,xspace}
\usepackage{epstopdf}
\usepackage{lineno}

\newcommand{\pp}{\ensuremath{\mathrm {p\kern-0.05em p}}\xspace}
\newcommand{\pbpb}{\ensuremath{\mbox{Pb--Pb}}\xspace}
\newcommand{\ppb}{\ensuremath{\mbox{p--Pb}}\xspace}

\newcommand{\sqrts}{\ensuremath{\sqrt{s}}\xspace}
\newcommand{\sqrtsnn}{\ensuremath{\sqrt{s_{\mathrm{NN}}}}\xspace}
\newcommand{\sqrtse}[2][TeV]{$\sqrts = #2\,\mathrm{#1}$}
\newcommand{\sqrtsnne}[2][TeV]{$\sqrtsnn = #2\,\mathrm{#1}$}
\newcommand{\pthere}{\ensuremath{p_{\mathrm{T}}}\xspace}
\newcommand{\Bpt}{\ensuremath{\mathbf{\textit{p}}_{\mathrm{\mathbf{T}}}}\xspace}

\newcommand{\raa}{\ensuremath{R_{\mathrm{AA}}}\xspace}
\newcommand{\rpbpb}{\ensuremath{R_{\mathrm{PbPb}}}\xspace}
\newcommand{\rpa}{\ensuremath{R_{\mathrm{pA}}}\xspace}
\newcommand{\rppb}{\ensuremath{R_{\mathrm{pPb}}}\xspace}
\newcommand{\Braa}{\ensuremath{\mathbf{R_{\mathrm{AA}}}}\xspace}

\newcommand{\Brpa}{\ensuremath{\mathbf{R_{\mathrm{pA}}}}\xspace}

\newcommand{\taa}{\ensuremath{T_{\mathrm{AA}}}\xspace}

\newcommand{\mevc}{\ensuremath{\mathrm{MeV}\kern-0.05em/\kern-0.02em c}\xspace}
\newcommand{\gevc}{\ensuremath{\mathrm{GeV}\kern-0.05em/\kern-0.02em c}\xspace}
\newcommand{\gevcsq}{\ensuremath{\mathrm{GeV}\kern-0.05em/\kern-0.02em c^2}\xspace}

\newcommand{\la}{\ensuremath{\Lambda}\xspace}

\newcommand{\kzero}{\ensuremath{\rm K^0_s}\xspace}
\newcommand{\ltok}{\ensuremath{\Lambda/\rm K^0_s}\xspace}
\newcommand{\ptopi}{\ensuremath{\rm p/\pi}\xspace}
\newcommand{\ktopi}{\ensuremath{\rm K/\pi}\xspace}
\newcommand{\phitop}{\ensuremath{\phi/\rm p}\xspace}
\newcommand{\ptophi}{\ensuremath{\rm p/\phi}\xspace}
\newcommand{\Bltok}{\ensuremath{\mathbf{\boldsymbol{\Lambda}/\rm\mathbf{K^0_s}}}\xspace}
\newcommand{\Bptopi}{\ensuremath{\mathbf{\rm\mathbf{p}/\boldsymbol{\pi}}}\xspace}
\newcommand{\Bktopi}{\ensuremath{\mathbf{\rm\mathbf{K}/\boldsymbol{\pi}}}\xspace}

\newcommand{\Bptophi}{\ensuremath{\mathbf{\rm\mathbf{p}/\boldsymbol{\phi}}}\xspace}

\newcommand{\jpsi}{\ensuremath{{\rm J}\kern-0.02em/\kern-0.05em\psi}\xspace}

\begin{document}
\title{ALICE summary of light flavour results at intermediate and high \Bpt}

\author{Tuva Richert, on behalf of the ALICE collaboration}

\address{Lund University, Department of Physics, Div. of Particle Physics, Box 118, SE-221 00 Lund}

\ead{tuva.richert@hep.lu.se}

\begin{abstract}
The ALICE experiment has unique capabilities for particle identification at mid rapidity over a wide range of transverse momenta (\pthere), making it an ideal tool for comprehensive measurements of hadrons such as charged pions, kaons, and protons as well as \la, \kzero and $\phi$. The transverse momentum distributions and nuclear modification factors, \rppb and \rpbpb, of these hadrons measured in p--Pb and Pb--Pb collisions are presented. Baryon-to-meson ratios exhibit a multiplicity-dependent enhancement at intermediate transverse momenta for both p--Pb and Pb--Pb collisions, while no significant dynamics is observed in the ratios at larger transverse momenta. Finally, measurements of identified particle ratios in association with high-\pthere particles as well as within reconstructed jets are presented. 
\end{abstract}

\section{Introduction}

During LHC Run-1 the ALICE detector has recorded \pp, \ppb, and \pbpb collisions at different center of mass energies. Heavy-ion collisions at ultra relativistic energies are expected to produce a QCD matter where the quarks and gluons are in a deconfined state. Measurements of the production of hadrons in \pbpb collisions at intermediate and high \pthere, relative to \pp collisions,
provide information about the dynamics of this matter. In the context of light-flavour production, the focus for the \pbpb results is on parton energy loss --- expected to lead to a modification of energetic jets (jet quenching) --- and possibly modified fragmentation due to the hot and dense QCD medium. 

The excellent tracking and particle identification capabilities of the ALICE experiment, in particular its large time projection chamber, makes it possible to investigate the spectra of baryons and mesons.
The results are presented in terms of particle ratios and nuclear modification factors, \raa and \rpa.



\vspace{0.3cm}
\section{\Braa and \Brpa for charged hadrons}

The nuclear modification factor is defined as the ratio of the particle yield in \pbpb to that in \pp collisions scaled by the number of binary nucleon-nucleon collisions
\begin{equation}
R_{\mathrm{AA}}(p_{\mathrm{T}}) = \frac{
  \mathrm{d}^2 N_{\mathrm{AA}} / \mathrm{d} \eta \mathrm{d}p_{\mathrm{T}}
  }{
    \langle T_{\mathrm{AA}} \rangle \mathrm{d}^2 \sigma_{\mathrm{pp}}/\mathrm{d}\eta \mathrm{d}p_{\mathrm{T}}
  }
\end{equation}
where $\mathrm{d}^2 N_{\mathrm{AA}} / \mathrm{d} \eta \mathrm{d}p_{\mathrm{T}}$ is the differential particle yield in \pbpb collisions, $\mathrm{d}^2 \sigma_{\mathrm{pp}}/\mathrm{d}\eta \mathrm{d}p_{\mathrm{T}}$ is the invariant cross section for particle production in inelastic \pp collisions, and $\langle T_{\mathrm{AA}} \rangle$ is the average nuclear overlap function \cite{ppb}. In the absence of nuclear modifications \raa is unity for hard processes which are expected to exhibit binary collision scaling.

\begin{figure}[htbp]
  \centering
  \includegraphics[keepaspectratio, width=0.65\textwidth]{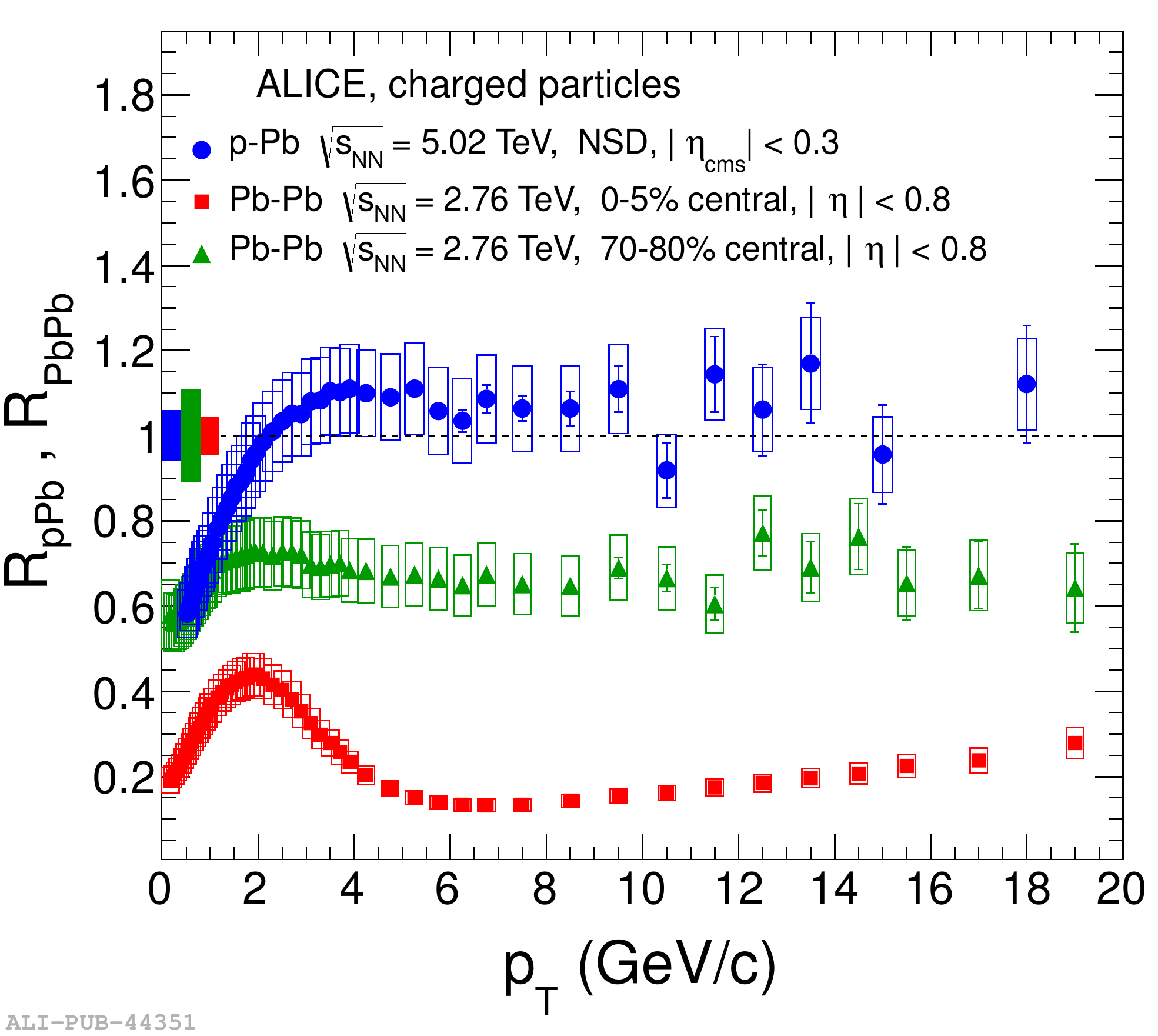}
  \caption{Nuclear modification factors, \rpbpb and \rppb, of charged particles as a function of transverse momentum for 0-5\% central and 70-80\% peripheral \pbpb at \sqrtsnne{2.76} collisions, and central \ppb at \sqrtsnne{5.02} collisions \cite{ppb}.}
  \label{fig:nonID}
\vspace{-0.2cm}
\end{figure}

The nuclear modification factor presented in Fig.~\ref{fig:nonID}, shows that the shape of the invariant yield for peripheral \pbpb collisions at \mbox{\sqrtsnne{2.76}} is similar to those observed in pp collisions due to the flatness of the \rpbpb, while a strong suppression of charged hadron production at high \pthere is observed for central collisions. To establish whether the initial state of the colliding nuclei plays a role in the observed suppression, also the nuclear modification factor in \ppb for charged particles is shown in the same figure. \rppb is consistent with unity for \pthere$>$ 2 \gevc, 
and hence the suppression in \pbpb collisions is not due to any initial state effects, but to final nuclear matter effects, such as jet quenching in the hot QCD medium. 

\section{\Braa and \Brpa for identified hadrons}

\begin{figure}[htbp]
  \centering
  \includegraphics[keepaspectratio, width=0.57\columnwidth]{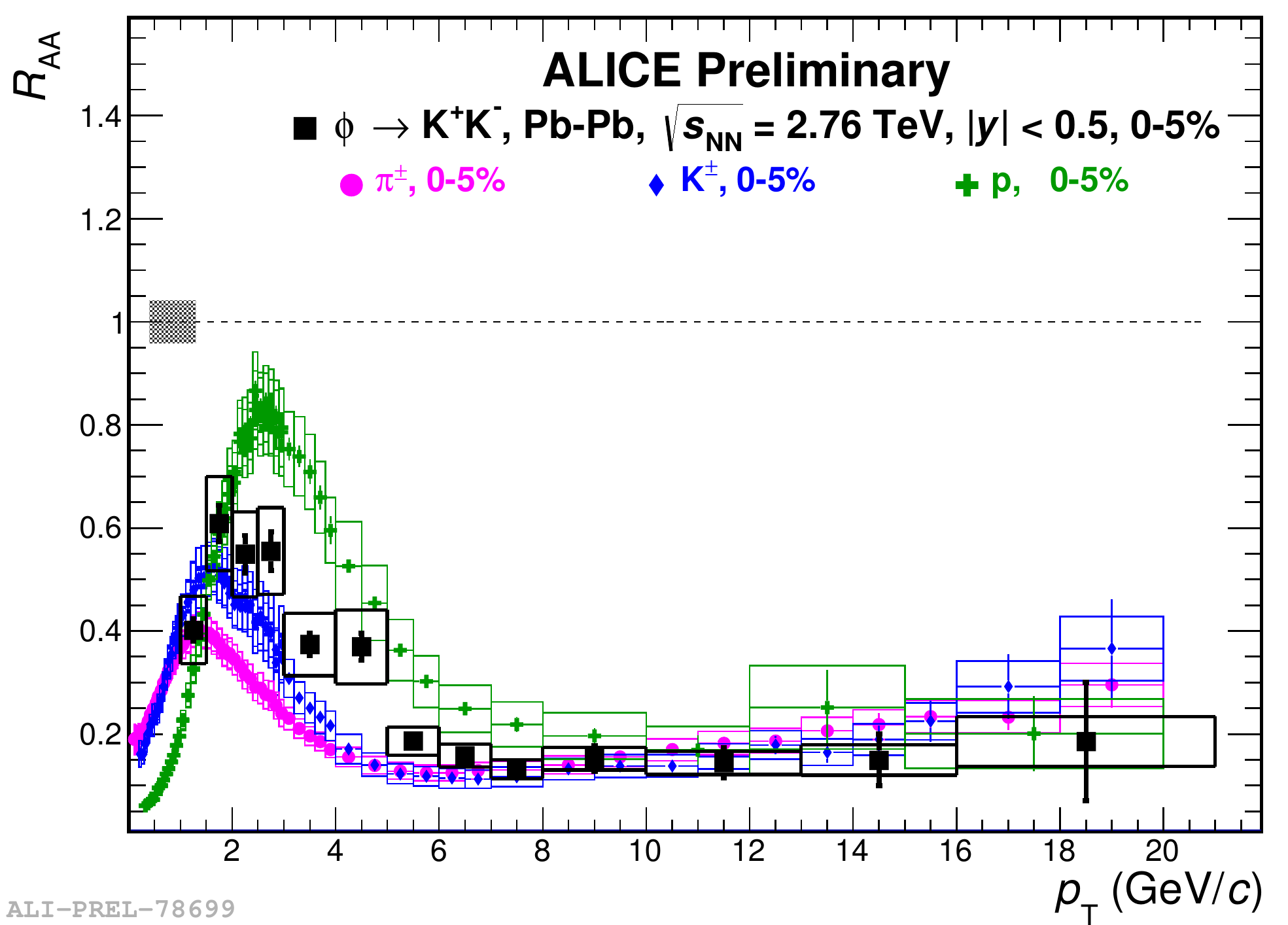}
  \caption{Nuclear modification factor, \raa, of charged $\pi$, K, p, and $\phi$ up to 20 \gevc in 0-5\% central \sqrtsnne{2.76} \pbpb collisions.}
  \label{fig:raa}
\end{figure}
\begin{figure}[htbp]
  \centering
  \includegraphics[keepaspectratio, height=0.36\columnwidth]{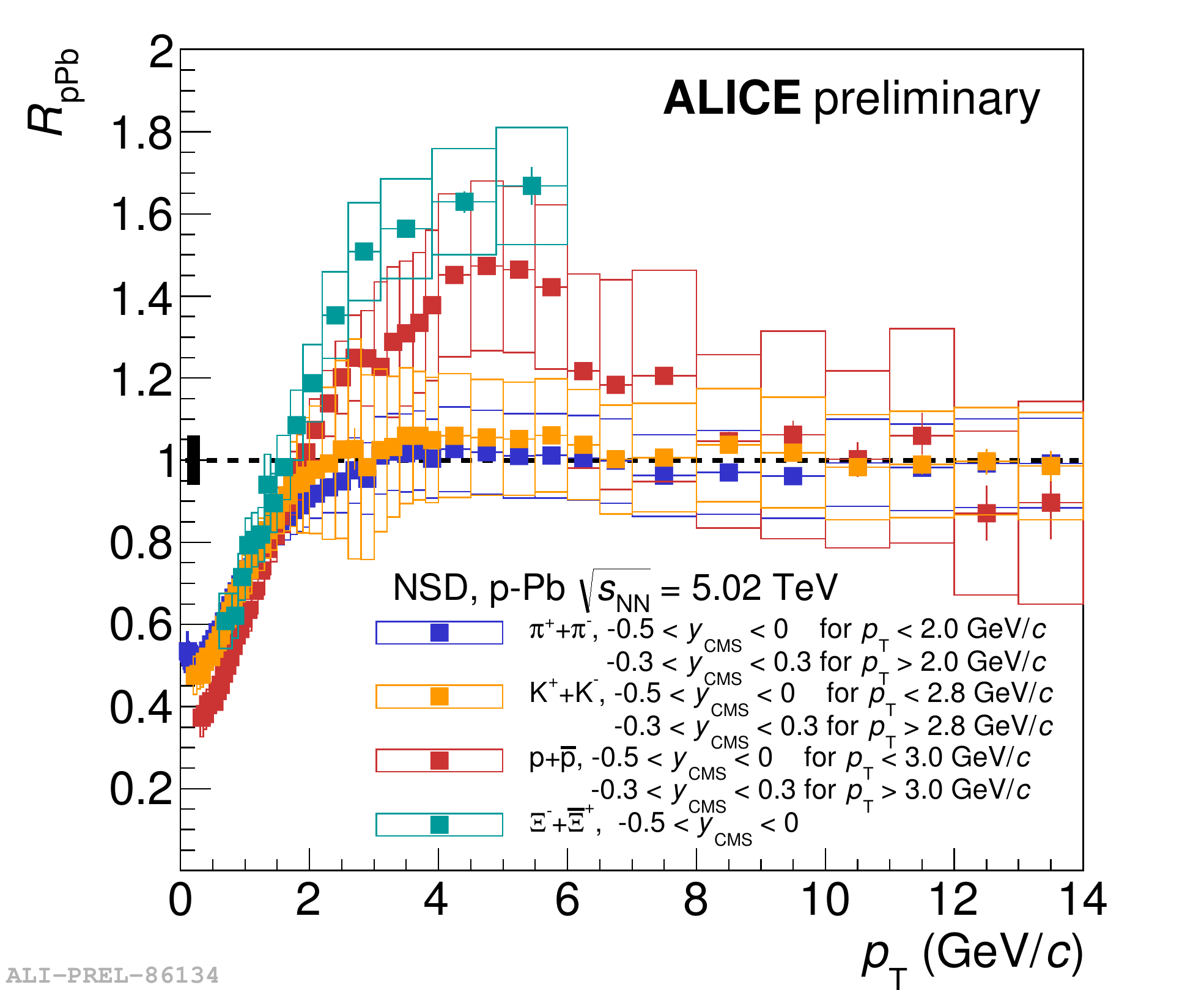}
  \includegraphics[keepaspectratio, height=0.35\columnwidth]{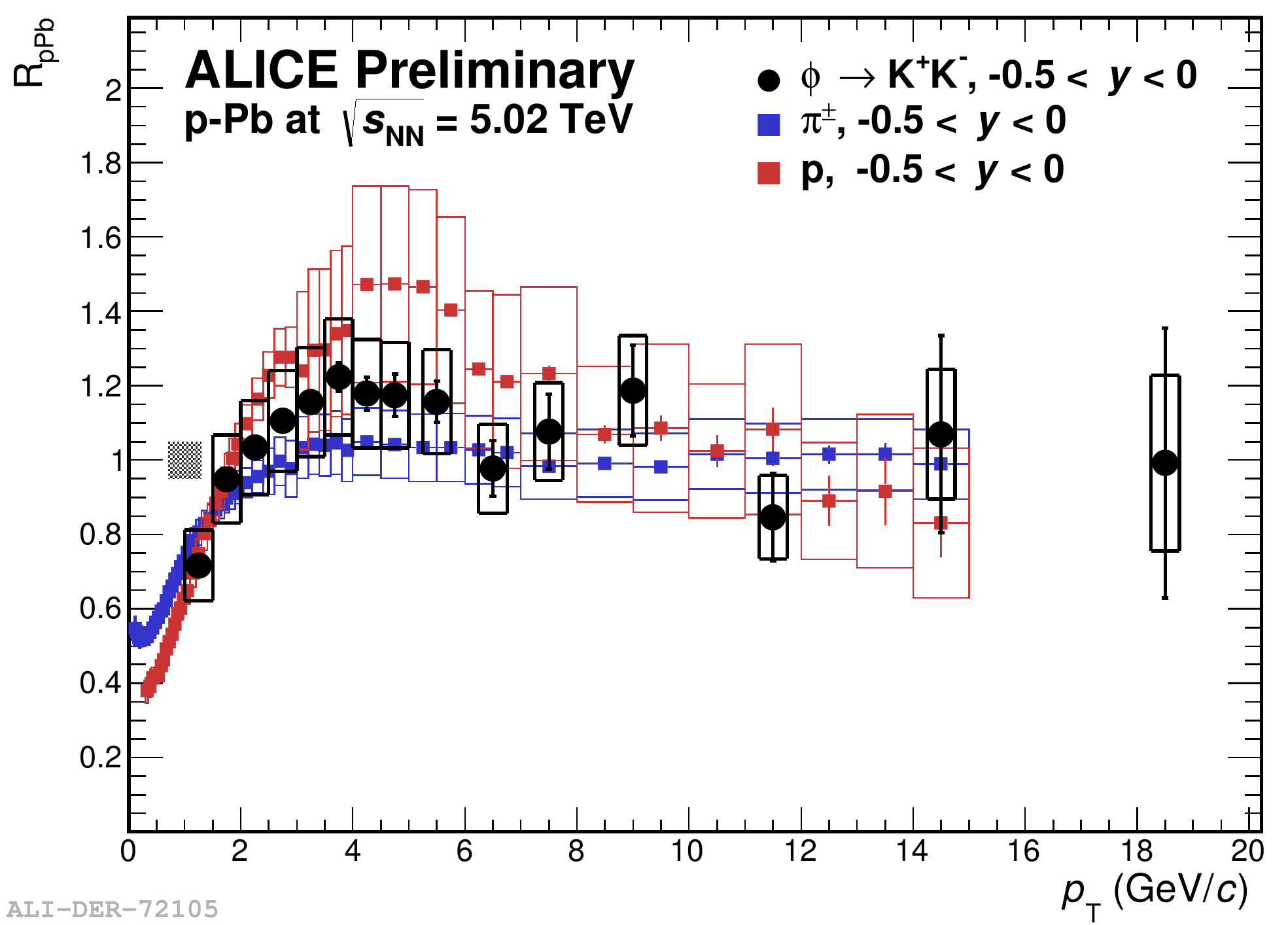}
  \caption{Nuclear modification factor, \rpa, for charged $\pi$, K, p, $\Xi$ (left) and $\phi$ (right) in central \sqrtsnne{5.02} \ppb collisions.}
  \label{fig:rpa}
\end{figure}

When constructing \raa for identified light flavour hadrons, we see in Fig.~\ref{fig:raa} that, within systematic and statistical uncertainties, they are equally suppressed at \mbox{\pthere$>$ 10 \gevc}. 
The large suppression is a sign of considerable energy loss, and the \rpa seen in Fig.~\ref{fig:rpa} establishes that this energy loss is predominantly due to the medium and not caused by initial state effects.

For the intermediate \pthere range, the protons are less suppressed than the kaons and pions, and a mass ordering is present in the suppression pattern. 
While the proton and $\phi$ modification factors exhibit rather distinct features, the \phitop ratio in central \pbpb is observed to be approximately constant as a function of \pthere, as discussed in Sec.~4, indicating that the differences in the \raa can be attributed to different \pp spectra shapes.


Looking at identified particles at intermediate \pthere in \ppb collisions instead, we see in Fig.~\ref{fig:rpa} an enhancement of $\Xi$ and p, while K and $\pi$ are consistent with \taa-scaled \pp values. 
Furthermore, there is a mass ordering among $\pi$, K, p, $\Xi$, but the $\phi$ does not fit into this pattern.

\section{\Bltok, \Bptopi and \Bktopi ratios in \pbpb collisions}



\begin{figure}[htbp]
  \centering
  \includegraphics[keepaspectratio, width=0.42\columnwidth]{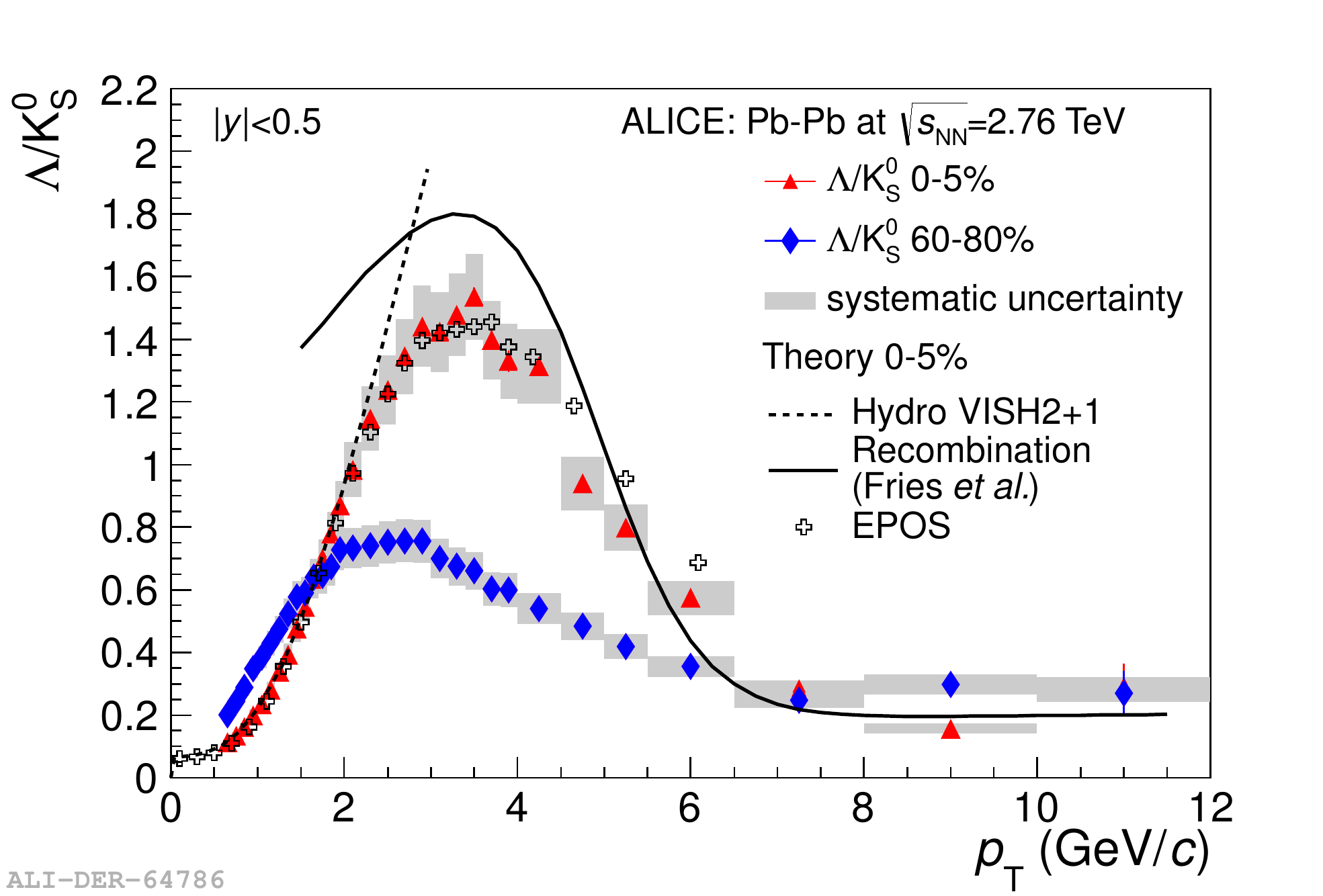}
  \includegraphics[keepaspectratio, width=0.56\columnwidth]{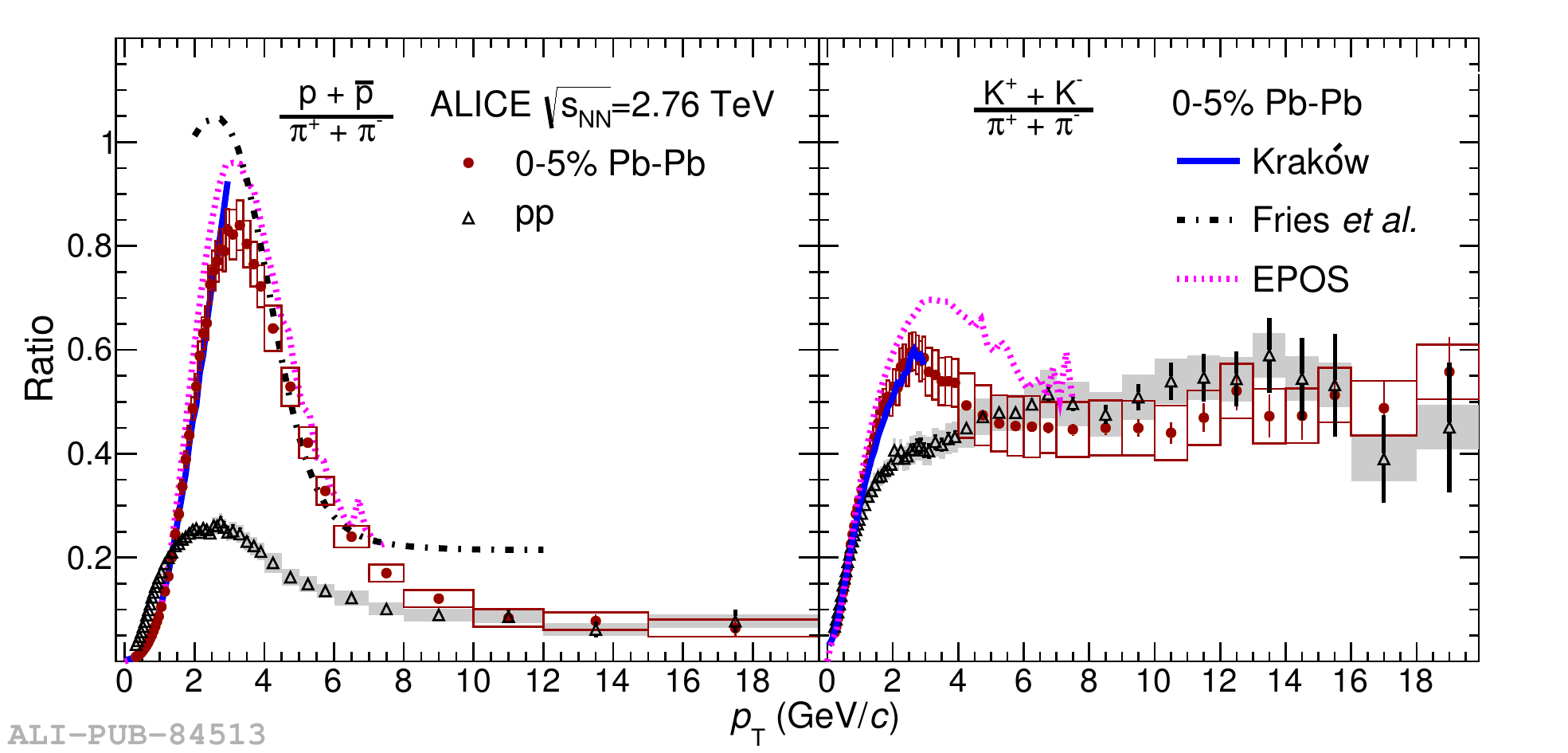}
  \caption{Left: \ltok particle ratio as a function of \pthere in central (0-5\%) and peripheral (\mbox{60-80\%}) \pbpb collisions at \sqrtsnne{2.76} compared to models in central (\mbox{0-5\%}) events \cite{strange}. Right: \ptopi and \ktopi ratio as a function of \pthere in central (\mbox{0-5\%}) \pbpb collisions at \sqrtsnne{2.76} compared to \pp collisions at \sqrtse{7} and models \cite{pikp}.}
  \label{fig:ltok}
\end{figure}

In \pbpb collisions, both \ltok and \ptopi (Fig.~\ref{fig:ltok} \cite{strange}, \cite{pikp}) in central and peripheral collisions are consistent with pp for \pthere$>$ 8 \gevc, indicating that the processes are dominated by vacuum-like fragmentation.

Looking in the intermediate \pthere range for \ltok, an enhancement is visible towards more central collisions, see Fig.~\ref{fig:ltok}, and a shift of the maximum position towards higher \pthere is observed: in the most peripheral collisions (60-80\% centrality) there is a maximum of about 0.55 at \mbox{\pthere$\sim$2 \gevc}, while the maximum value of the ratio for the most central collisions (0-5\% centrality) is about 1.6 at \pthere$\sim$3.2 \gevc. This shift is consistent with an increasing radial flow towards more central collisions.
The magnitude of these maxima increases by almost a factor of three between most peripheral and most central \pbpb collision. 
A hydrodynamical model such as VISH2+1 \cite{krakow} is able to describe the rise at low \pthere.
At higher \pthere, models with modified fragmentation (EPOS \cite{EPOS1}, \cite{EPOS2}) and coalescence of quarks (Recombination \cite{fries}) describe the shape qualitatively well, but overestimate the enhancement \cite{strange}.





Figure \ref{fig:ltok} also shows the \ptopi and \ktopi ratio up to \pthere= 20 \gevc for central events, both presenting an enhancement at intermediate \pthere, with the peak at \pthere= 3 \gevc. 
However, the baryon-to-meson ratio \ptopi presents a much more pronounced increase, reaching a value of about 0.9 at \pthere= 3 \gevc, compared to the two-meson ratio \ktopi.
As for the \ltok case, the ratios are in good agreement with hydrodynamical calculations (Krakow \cite{krakow}) for \mbox{\pthere$<$ 2 \gevc}, indicating that the rise of the peak can be described by the mass ordering induced by radial flow. 
At intermediate \pthere, around the maxima and up to \mbox{\pthere$\sim$8 \gevc}, the data are qualitatively described by the recombination model by Fries \textit{et al.} \cite{fries}, and the EPOS model \cite{EPOS1}, \cite{EPOS2}, but these models also overestimate the maximum values.

\vspace{0.2cm} 
\section{\Bptophi in \pbpb collisions}



\begin{figure}
   \centering
   \includegraphics[keepaspectratio, width=0.55\columnwidth]{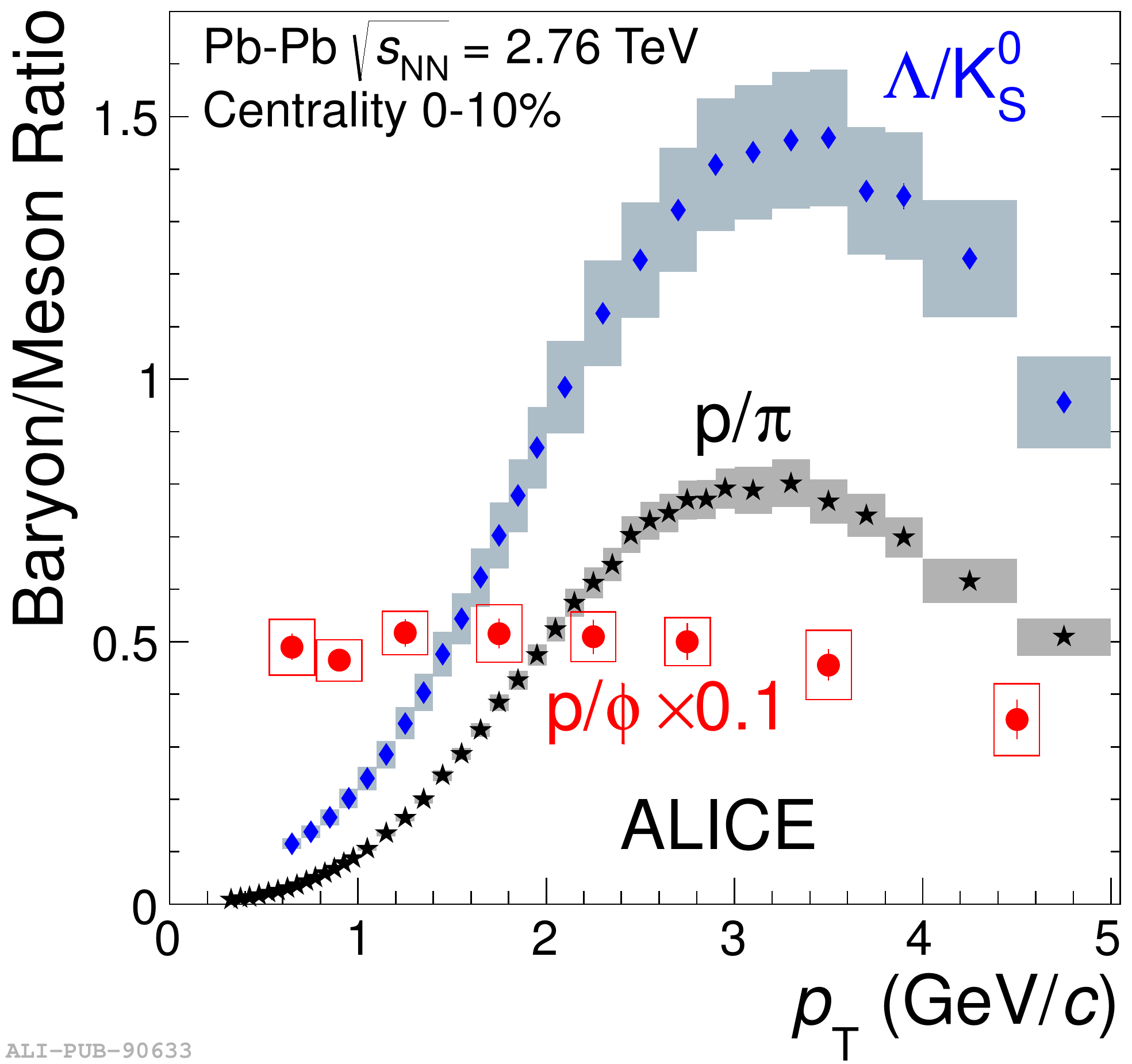}
   \caption{Ratio \ptophi (scaled with 0.1 for visibility) as a function of \pthere up to 5 \gevc for 0-10\% central \pbpb collisions at \sqrtsnne{2.76}, compared to \ltok and \ptopi in the same centrality bin. Figure taken from \cite{phi}.}
   \label{fig:phitop}
 \end{figure}

 To further investigate the main driving parameter in the spectral shape, we study a baryon-to-meson ratio in which the baryon and meson are of similar mass, namely the \ptophi ratio. In Fig.~\ref{fig:phitop} \cite{phi} the \ptophi ratio is shown as a function of \pthere, and it is observed that in \textit{central} \pbpb collisions there is a very small difference in their \pthere distributions, i.e. no baryon-meson difference is present. This indicates that the hadron mass determines the spectral shape. 

 \vspace{0.2cm} 
 \section{Particle ratios in \ppb collisions}

\begin{figure}[htbp]
  \centering
  \vspace{-0.2cm}
  \includegraphics[keepaspectratio, width=0.47\columnwidth]{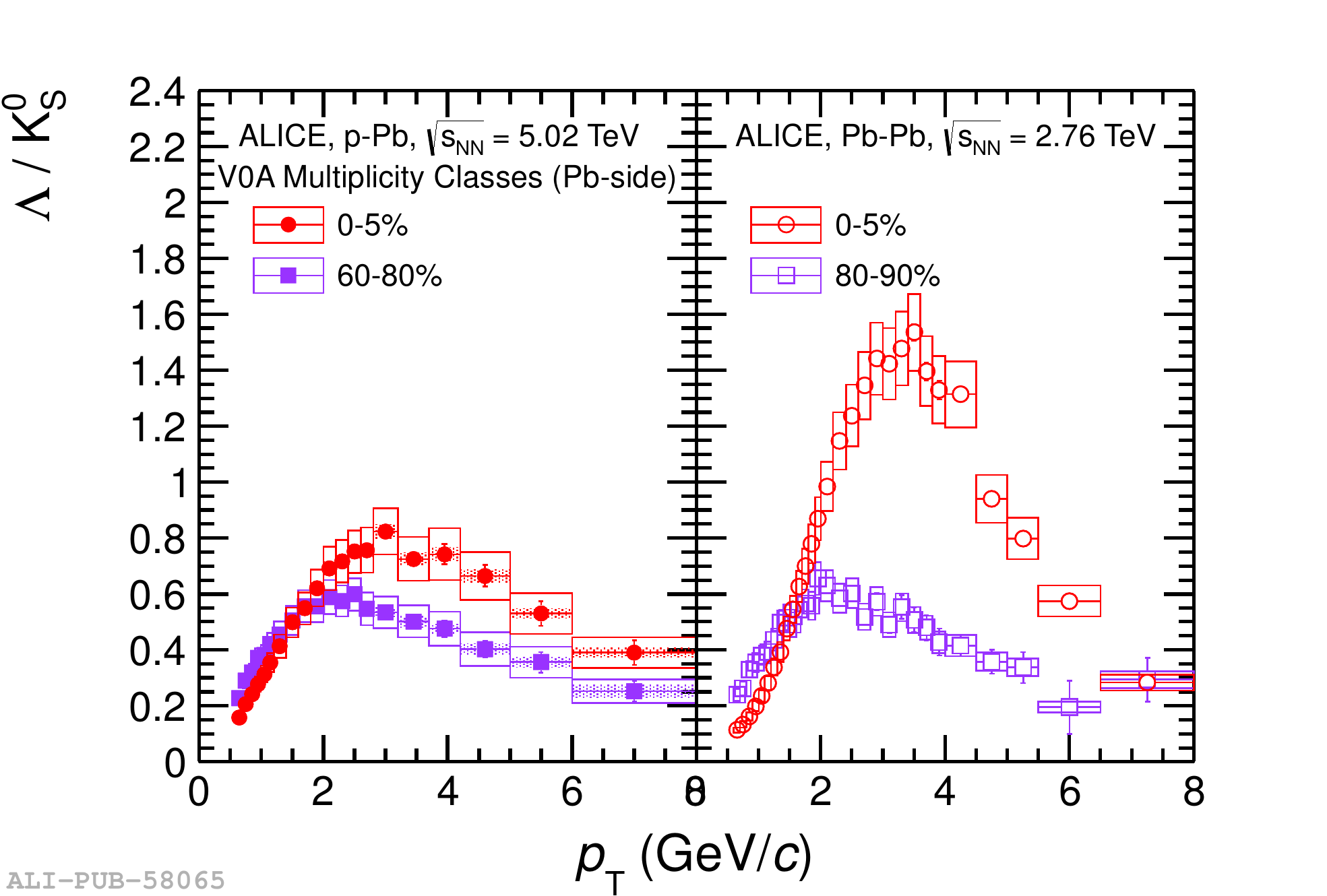}
  \includegraphics[keepaspectratio, width=0.47\columnwidth]{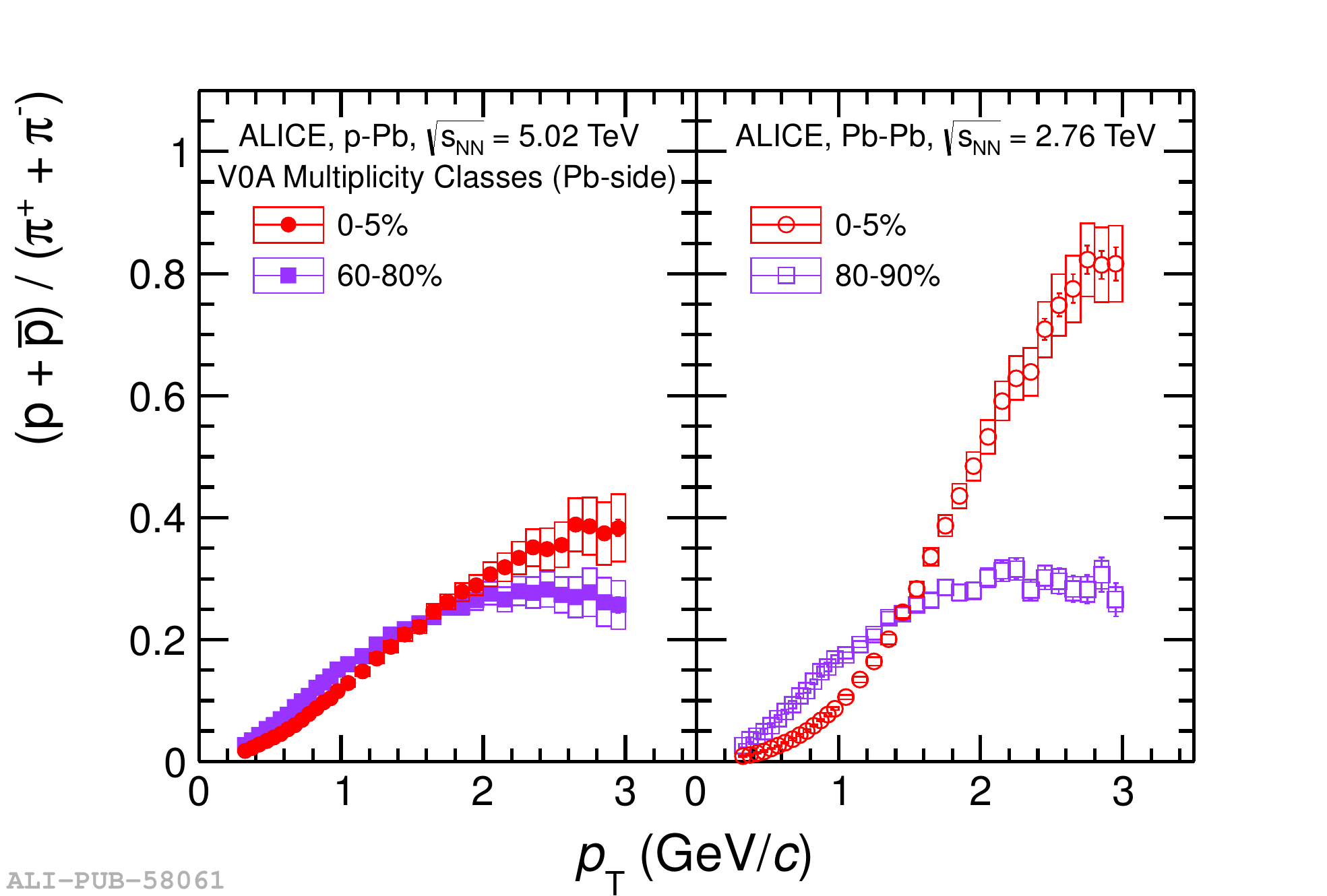}
  \vspace{0cm}
  \caption{\ltok (left) and \ptopi (right) in \sqrtsnne{5.02} \ppb collisions (left panel) as a function of \pthere in one central (0-5\%) and one peripheral (60-80\%) multiplicity bin. The ratios are compared to results in \pbpb collisions (right panels) where central multiplicity bin is 0-5\%, and the peripheral multiplicity bin is 80-90\%. Figures from \cite{particle_ppb}.}
  \label{fig:particle_ppb}
\end{figure}

Interestingly, the \ltok and \ptopi in \ppb collisions show the same qualitative behavior as in \pbpb collisions: a multiplicity dependent baryon-to-meson enhancement at intermediate \mbox{\pthere$\sim$3 \gevc} is seen in Fig.~\ref{fig:particle_ppb} \cite{particle_ppb} for two different multiplicity event classes. The results show that \ppb presents features that are similar to \pbpb phenomenology,
even though the magnitude of the enhancement in \ppb is significantly different to the one observed in \pbpb. The maximum of the \ptopi ratio reaches 0.8 in central \pbpb collisions, but only 0.4 in the highest multiplicity \ppb events, and the \ltok maximum in central \pbpb is 1.5, while it is 0.8 in corresponding \ppb collisions. The highest multiplicity bin in \ppb collisions exhibits ratios of \ptopi and \ltok which have maxima close to the corresponding ratios in the 60-70\% centrality bin in \pbpb collisions, but differ somewhat in shape at lower \pthere \cite{particle_ppb}.

\pagebreak
\section{The origin of the enhancement}

\begin{figure}[htbp]
  \centering
  \hspace{0.5cm}
  \includegraphics[keepaspectratio, width=0.6\columnwidth]{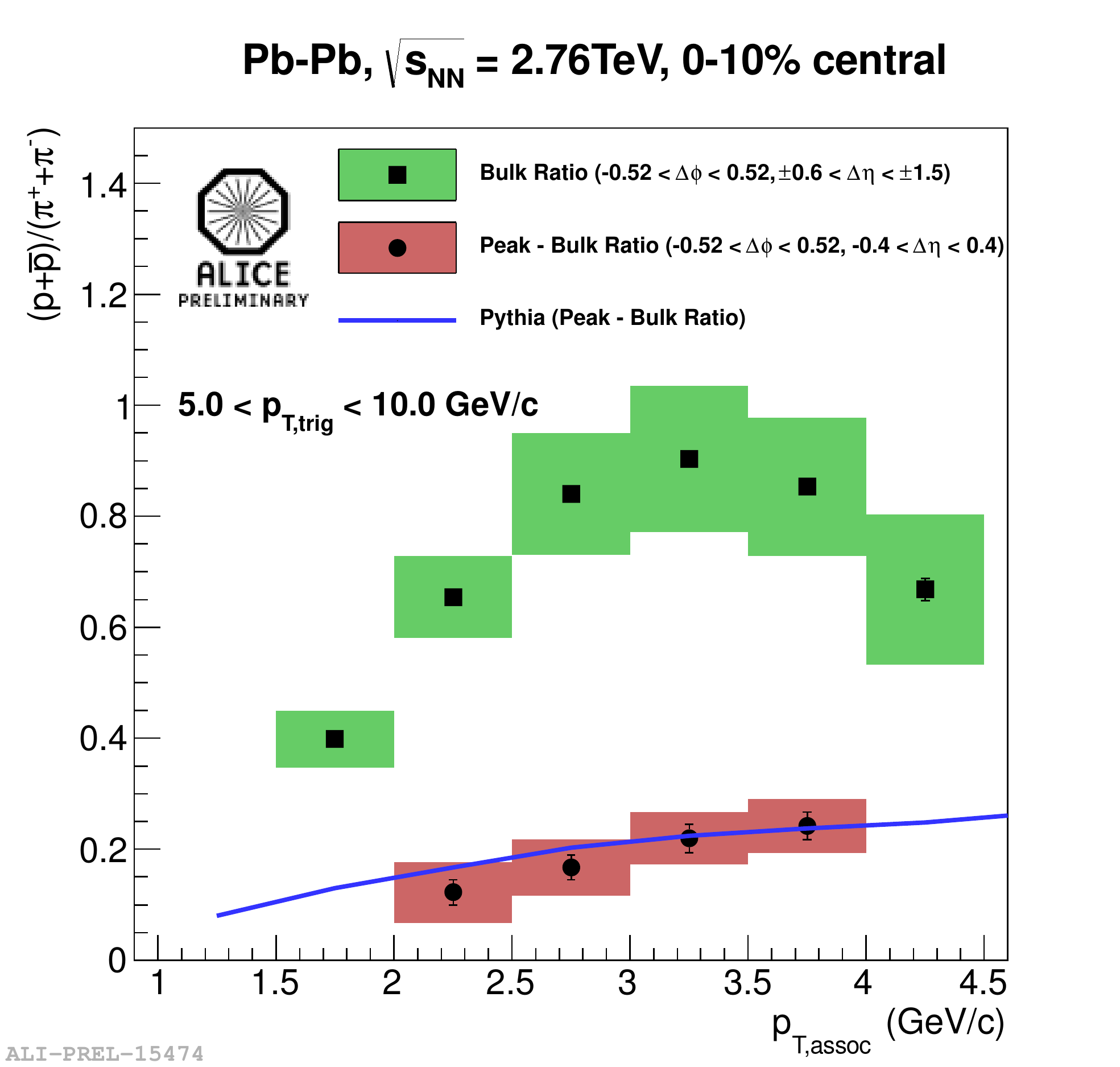}
  \hspace{-3cm}
  \includegraphics[keepaspectratio, width=0.56\columnwidth]{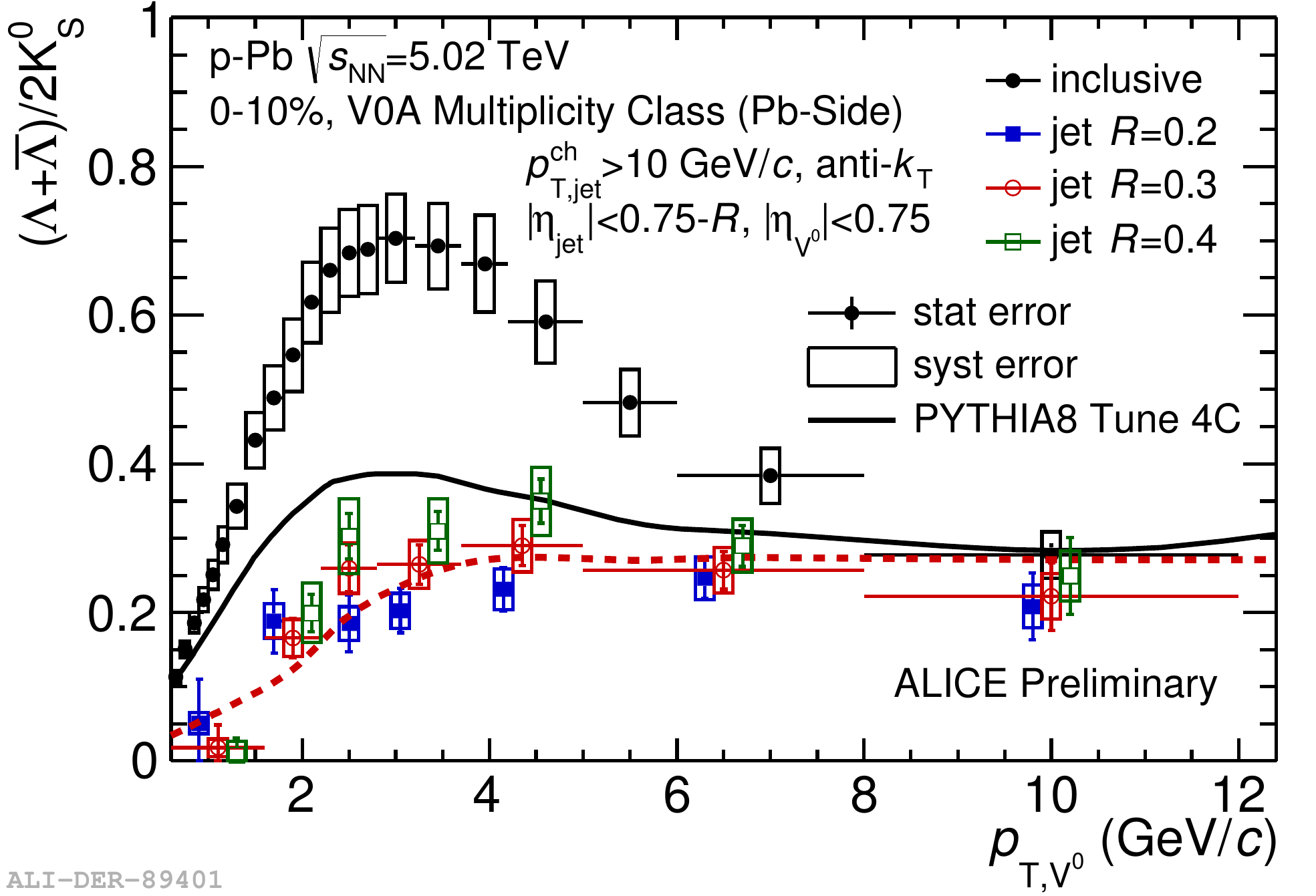}
  \caption{Top: \ptopi ratio as a function of associated particle \pthere, in bulk and peak-bulk for 0-10\% central \pbpb collisions at \sqrtsnne{2.76}, with a leading (trigger) particle \pthere between 5--10 \gevc \cite{crc}. Bottom: \ltok ratio in jet with different radii reconstructed with the \mbox{anti-$k_\mathrm{T}$} method compared to PYTHIA8 and the \pbpb inclusive ratio (full black circles) for 0-10\% central \ppb collisions at \sqrtsnne{5.02}.}
\label{fig:origin}
\end{figure}

One can investigate wether the origin of this enhancement is due to parton fragmentation (hard) or collective effects (soft) by a two-particle correlation study, where the particles produced in the underlying events, the \textit{bulk}, are separated from those which are associated with a high-\pthere trigger particle, representing a jet-like environment, or the \textit{peak} region. The peak is defined as a region around $(\Delta\eta,\Delta\varphi)=(0,0)$, and the bulk region around $(\Delta\eta,\Delta\varphi)=(\pm 1,0)$ where one, due to long range (in rapidity) azimuthal correlations, expects the flow structure of the underlying event to be the same as under the peak. To study the jet contribution, the bulk is subtracted from the peak region (in Fig.~\ref{fig:origin}: "Peak-Bulk"). In Fig.~\ref{fig:origin} (top) \cite{misha}, the \ptopi ratios in central \pbpb events are presented for bulk and for peak-bulk event selections, and it is seen that the enhancement is a bulk effect and nt present in jet events.

In the \ppb study, charged particle jets are reconstructed on an event-by-event basis using an anti-$k_\mathrm{T}$ algorithm with resolution parameter $R$ = 0.2, 0.3, or 0.4 and requiring one charged track with \pthere$>$ 10 \gevc. The \la and \kzero yields are measured within the jet cone and corrected
for the underlying event before the ratio is taken. When the ratio is
compared to the inclusive ratio, the same conclusion as for the \ptopi ratio can be drawn: that the baryon-to-meson enhancement originates from the bulk, and is not present in the jet structure. 

%
%



\pagebreak
\section{Conclusions}

At high \pthere we observe a suppression of identified particle production due to parton energy loss. The same suppression is seen for all light quark systems created in \pbpb collisions, which suggests that the chemical composition of leading particles from jets in the medium is similar to jets produced in vacuum. No suppression in \ppb collisions is seen, indicating that the suppression observed in \pbpb is a final-state hot-matter effect. 

At intermediate \pthere, the particle ratios show a baryon-to-meson enhancement which in \pbpb is understood in the coalescence and/or hydrodynamic flow picture. 
In \ppb collisions we see similar features, but less pronounced, as in \pbpb. By separating the underlying events from the jet-like structures, we note that the baryon-to-meson enhancement seems to be an effect arising in the underlying events in both \pbpb and \ppb collisions, while the jet-like contributions appear to be unmodified.

\end{document}